\documentclass[11pt]{article}

\usepackage{hyperref}
\hypersetup{
    colorlinks=true,
    linkcolor=blue,
    filecolor=magenta,      
    urlcolor=blue,
}

  \usepackage{geometry}
\usepackage{amsmath,amssymb,amsthm}
\usepackage{graphics,epsfig,calc}
\usepackage{amsmath}

\usepackage{latexsym,epsfig,bm,amssymb}
\usepackage{xcolor}
\usepackage{amsthm,mathrsfs}

\usepackage{mathptmx}

\DeclareSymbolFont{AMSb}{U}{msb}m{n}
\DeclareSymbolFontAlphabet{\mathbb}{AMSb}

 \newcommand{\beqn}{\begin{eqnarray}}
 \newcommand{\eeqn}{\end{eqnarray}}
 \newcommand{\be}{\begin{equation}}
 \newcommand{\ee}{\end{equation}}
  
  \newcommand{\bcor}{\begin{corollary}}
 \newcommand{\ecor}{\end{corollary}}
 
 \newcommand{\bpr}{\begin{proof}}
 \newcommand{\epr}{\end{proof}}
 
 \newcommand{\bex}{\begin{example}}
 \newcommand{\eex}{\end{example}}

 \newcommand{\ba}{\begin{array}}
 \newcommand{\ea}{\end{array}}

 \newcommand{\pa}{\partial}
 
  \newcommand{\ci}{\cite}
 
 \newcommand{\la}{\label}
 \newcommand{\rIm}{{\rm Im\5}}
 \newcommand{\rRe}{{\rm Re\5}}
 \newcommand{\fr}{\frac}

\newcommand{\ov}{\overline}

\newcommand{\ti}{\tilde}

\newcommand{\cF}{{\cal F}}

\newcommand{\ve}{\varepsilon}
\newcommand{\vp}{\varphi}
\newcommand{\De}{\Delta}

\newcommand{\al}{\alpha}

\newcommand{\ga}{\gamma}
\newcommand{\Ga}{\Gamma}
\newcommand{\vka}{\varkappa}

\newcommand{\om}{\omega}
\newcommand{\Om}{\Omega}
\newcommand{\na}{\nabla}

\newcommand{\5}{{\hspace{0.5mm}}}

\newcommand{\R}{\mathbb{R}}

\newcommand{\C}{\mathbb{C}}

\newtheorem{theorem}{Theorem}[section]

\newtheorem{defin}[theorem]{Definition}

\newtheorem{lemma}[theorem]{Lemma}
\newtheorem{remark}[theorem]{Remark}
\newtheorem{remarks}[theorem]{Remarks}
\newtheorem{corollary}[theorem]{Corollary}
\newtheorem{pro}[theorem]{Proposition}
\newtheorem{example}[theorem]{Example}

\newcommand{\bp}{\begin{pro}}
\newcommand{\ep}{\end{pro}}
\newcommand{\bl}{\begin{lemma}}
\newcommand{\el}{\end{lemma}}
\newcommand{\bc}{\begin{corollary}}
\newcommand{\ec}{\end{corollary}}

\newcommand{\bd}{\begin{defin}}
\newcommand{\ed}{\end{defin}}
\newcommand{\br}{\begin{remark}}
\newcommand{\er}{\end{remark}}
\newcommand{\brs}{\begin{remarks}}
\newcommand{\ers}{\end{remarks}}

\newcommand{\bce}{\begin{center}}
\newcommand{\ece}{\end{center}}

\begin{document}
\begin{center}

{\huge On Malyshev's method of automorphic functions 
\medskip\\

in diffraction by wedges}
\medskip\\

\bigskip\smallskip

\hfill{\it\large Dedicated to the memory of Vadim Malyshev}
 
\bigskip\smallskip

 {\Large A.I. Komech
  }
  \smallskip\\
{\it   Dobrushin Lab., IITP RAS, Moscow\\
Department Mechanics-Mathematics, Moscow State University}
\\
akomech@iitp.ru
 \bigskip\\
{\Large A.E. Merzon}\footnote{The research supported by  CONACYT-M\'exico and CIC-UMSNH, M\'exico.}
\\
{\it  Institute of Physics and Mathematics\\
 University of Michoacan de San Nicolas de Hidalgo, Morelia, Mexico}
 \\
 anatoli.merzon@umich.mx

\end{center}


\begin{abstract}
We describe Malyshev's method of automorphic functions in application
to boundary value problems in angles and to diffraction by wedges.
We give a consize survey  of related results of A. Sommerfeld, S.L. Sobolev,
J.B. Keller, G.E. Shilov and others.

  \end{abstract}


  \smallskip
  

\tableofcontents

\section{Introduction}
\smallskip

Vadim Malyshev was a very talented and versatile mathematician. He owns significant results in the field of probability theory and Gibbs fields,
Markov processes and Euclidean quantum field theory.
He also possessed outstanding organizational skills, in particular, he founded the successful and respected mathematical journal ``Markov Processes and Related Fields".

In 1970, V. Malyshev  invented  the method of automorphic functions \ci{M1970},
 and applied to random walks on the lattice
in the quarter of plane. Later on, he applied the method
to queueing systems and analytic combinatorics \ci{FIM2017}. In 1972--2022, the method  was extended to boundary value problems 
 for partial differential equations in angles \ci{K1973,KMZ2002} and to  diffraction by wedges \ci{KM2019}. 
 The main steps of Malyshev's method are as follows:
    \medskip
    
 I. Undetermined algebraic equation on the Riemann surface and analytic continuation.
   \smallskip
   
 II. Elimination of one unknown function using covering automorphisms.    
 \smallskip
 
   III.  The reduction to the Riemann--Hilbert problem.
  \smallskip

 Malyshev's method played  the crucial role in 
 the progress in the theory of diffraction by wedges 
 with general boundary conditions
 since 1972.
 The problem was stated by M.I.Vishik in the Summer  of 1967.
   In 1969--1971, 
  one of the authors (AK) tried to solve this problem while
  preparing 
  his PhD Thesis.
      As the result of these three-year efforts, 
  the problem has been reduced to an  undetermined algebraic equation on the Riemann surface \ci{K1974}, though next steps remained obscure. Fortunately,
  at the end of 1971,
AK received the impetus from his friend
Alexander Shnirelman who noticed something similar in  Malyshev's  book \ci{M1970}, which he had recently reviewed by request of M.I. Vishik.
AK did not understand this book completely, but discovered 
 two pages  which could have contained a creative idea.
The book, opened on these pages, lied on  his desk for about two or three months, when AK  pinned down two lines with the key idea
of automorphicity. The remaining work took about six months...

The extension of the research  to diffraction problems 
was done in  an  intensive collaboration  of both authors, 
and took about 50 years. The main results of the collaboration
were the {\it limiting absorption principle} \ci{M1973,M1977}, proof of the completeness of Ursell's trapping modes \cite {{KMZ1996}} ,
the extension to the 
nonconvex angles \ci{KM1992,KM2019}, and
the Sommerfeld representation \ci{KMM2005}. 
Moreover, our general methods \ci{KM2019}
allowed us to reproduce the formulas obtained by 
Sommerfeld, Sobolev and Keller \ci{KM2015,KMNMV2018,MKMV2015}.
The identifications justify these formulas as the 
{\it limiting amplitudes} in diffraction.

In  the present, we give a concise survey of the development
 of
Malyshev's method of  automorphic functions  since 1972
in the context of  i) 
boundary value problems in angles for elliptic partial differential equations,
and   ii)   theory of stationary and time-dependent diffraction by wedges.
We focus on principal ideas omitting nonessential technical details. 
All the details can be found in \ci{KM2019}.

\section{Diffraction by wedges and radar/sonar detection}
\begin{figure}[htbp]
		\begin{center}
			\includegraphics[width=1.1\columnwidth]{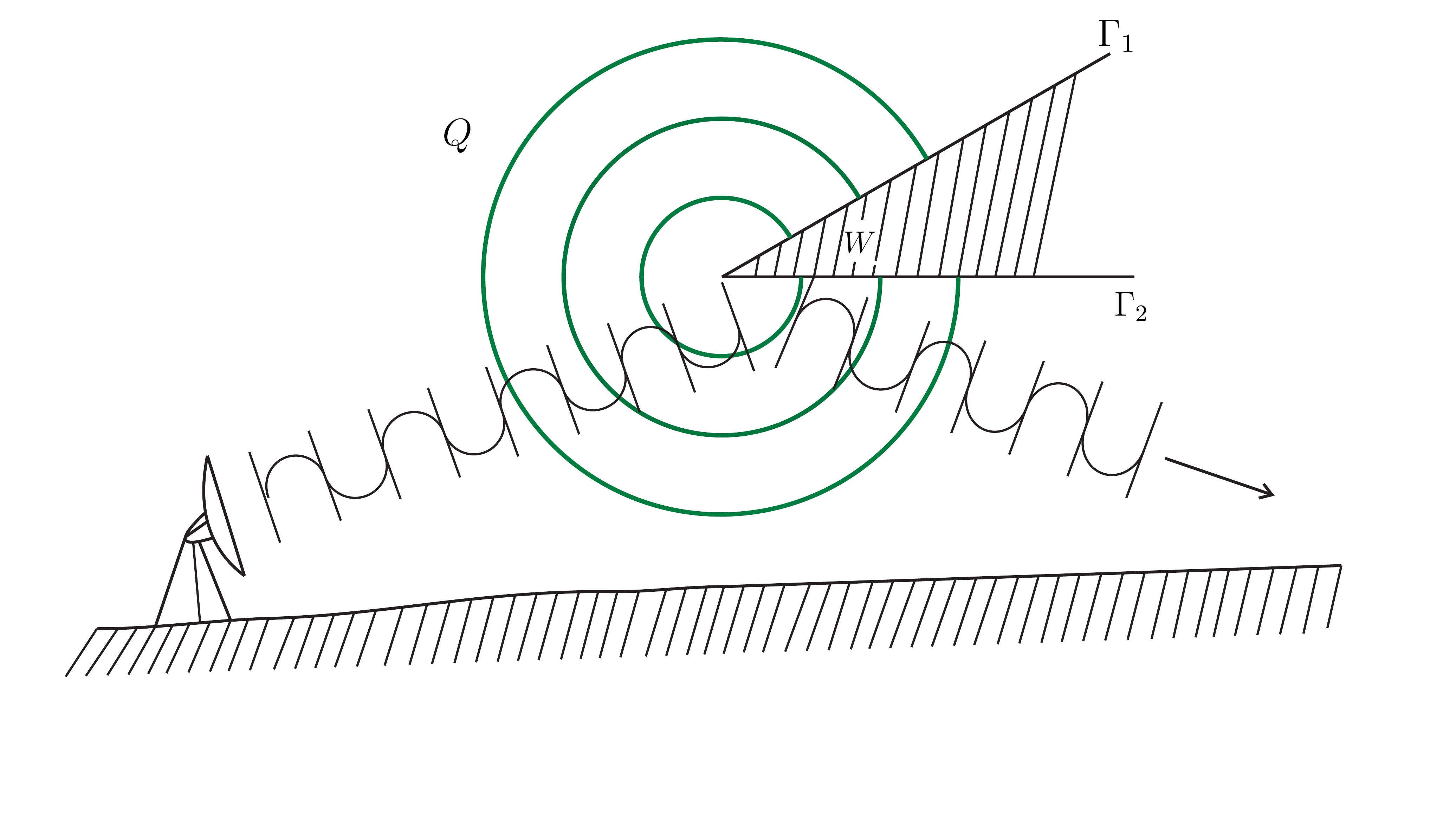}
			\caption{Incident, reflected, and diffracted waves (the latter in green color).}
			\label{fig1}
		\end{center}
	\end{figure}

The radar or sonar emits the incident wave,  which generates the reflected and diffracted waves (the latter in green color)
 as  shown in Fig. \ref{fig1}.
Here  $W$ denotes a conducting wedge (for example, the edge of  an 
	airplane wing), and Q$=\R^2\setminus W$ is  an angle of magnitude $\Phi$.
	The incident wave reaches the wedge and generates the reflected and  diffracted waves. The diffracted wave is defined as the total wave minus the incident and reflected
	waves.
	
	The reflected wave is defined by geometric optics, and  is absorbed by the ground. On the other hand, the diffracted wave spreads in all directions, and {\bf only this part of radiation} 
	 returns to the radar  which  allows to detect the airplane location.

\qquad

\section{Stationary diffraction and boundary value problems in angles}

The stationary diffraction by wedge is described by the 
 boundary value problem for the Helmholz equation in  an  angle $Q\subset\R^2$
of magnitude $\Phi\in(0,2\pi]$:
\be\la{Helm}
\left\{\ba{rcll}
(\De+\om^2) u(x)&=&0,& x\in Q
\\\\
  B_lu(x)&=&f_l(x),&x\in\Ga_l,\quad l=1,2
  \ea\right|,
  \ee
  where $\Ga_1$ and $\Ga_2$ denote the sides of the  angle, 
  the functions $f_l$ are defined by the incident wave, and 
  $\om\in\R$ is its frequency, see Fig. \ref{fig1} and (\ref{nsd3}).
  The operators $B_l$ in the boundary conditions 
  correspond to the  material   properties of the wedge (conductor, insulator, ferromagnetic, etc).

  The relation of stationary problem (\ref{Helm}) to time-dependent diffraction is highly nontrivial. The key issue is that for $\om\in\R$, the problem admits an infinite number
  of linearly independent solutions. We discuss this issue  in detail
 in Section \ref{std}.

The stationary diffraction  problem (\ref{Helm}) with the
Dirichlet and Neumann boundary conditions ($B_l=1$ or $B_l=\fr{\pa}{\pa n}$)
was solved in 1896--1912 
for $\Phi=2\pi$, 
  by A. Sommerfeld \ci{S1896}--\ci{S1954} (the detailed exposition and comments can be found in \ci{NZS2004}).
  The extension to 
 all $\Phi\in(0,2\pi)$ was obtained  in 1920 by H.S. Carslaw  \ci{C1920}, in 1932--1937 by V.I. Smirnov and  S.L. Sobolev \ci{SS1932,S1934,S1935,S1937,S1958}, and in 1951 by J.B. Keller and  A. Blank \ci{KB1951}.
  In 1958, 
G.D. Malujinetz solved the problem for all $\Phi\in(0,2\pi)$ with the impedance (Leontovich) boundary condition $\fr{\pa u(x)}{\pa n}+ib_l u(x)=f_l(x),\,\, x\in\Ga_l$;
see \ci{M1958,M1959}.
The detailed exposition of all these results can be found in \ci{BLG2007} and \ci{KM2019}.
\smallskip

The {\it  mixed  boundary value problems} of type 
\be\la{AB}
\left\{\ba{rcll}
A u(x)&=&0,& x\in Q
\\\\
 B_l u(x)&=&f_l(x),& x\in\Ga_l,\quad l=1,2
  \ea\right|,\qquad A=\sum_{|\al|\le m} a_\al\pa^\al,\quad B_l=\sum_{|\al|\le n_l} b_{l\al}\pa^\al
  \ee
 were considered in 1958 by S.L. Sobolev \ci{S1958}  and in 1960--1961 by G.E. Shilov \ci{S1960,S1961} in the quadrant $x_1>0$, $x_2>0$ for the case of hyperbolic operator $A$ in the variable $x_2$ and with the Cauchy initial conditions at $x_2=0$.
\smallskip

For strongly-elliptic  second order operators $A$ and general differential boundary operators $B_l$,
the problem (\ref{AB}) was
solved in 1972 in 
 {\bf convex angles} $Q$ of magnitude
$\Phi\in(0,\pi)$, 
see \ci{K1973,K1974}.   Strong ellipticity means
that
\be\la{sel}
|\hat A(z)|\ge \vka(|z|^2+1),\qquad z\in\R^2,
\ee
where  {\bf the symbol} $\hat A(z):=\sum_{|\al|\le 2} a_\al(-iz)^\al$ and
$\vka>0$.
 In particular, the operator   $A=-\De+1$ with the symbol $\hat A(z)=z^2+1$ is strongly elliptic, and also the Helmholtz operator 
$H=\De+\om^2$ 
from (\ref{Helm}) is strongly ellipltic  
for $\rIm\om\ne 0$. 
  The method
\ci{K1973,K1974} relies on the Malyshev ideas of automorphic functions \ci{M1970} which is presented in the next section.

The extension of this result to
nonconvex angles of magnitudes
$\Phi\in(\pi,2\pi)$, was done in 1992
 by the authors \ci{KM1992}.

Let us note that
the Helmholtz operator $A=\De+\om^2$ is not  strongly elliptic if $\om\in\R$ since
its symbol   has the form 
$\hat A(z)=-z^2+\om^2$.
  Problem (\ref{AB}) for the Helmholtz operator 
in {\bf convex angles}
was solved in 1972--1977
by A.E. Merzon \ci{M1973,M1977}, who proved  that
{\it for real $\om\in\R$},
the problem admits only a finite number of solutions satisfying
the {\bf limiting absorption principle}:
 \be\la{lab}
 u_\om(x)=\lim_{\ve\to 0+}u_{\om+i\ve}(x),\qquad x\in Q,
\ee
where $u_{\om+i\ve}$ denotes suitable  solution to (\ref{Helm}) with 
$\om+i\ve$ instead of $\om$.
\smallskip

The application of these results to time-dependent
diffraction by wedges was done in 2006--2019
by the authors
 \ci{KM2006,KM2007,KM2019}, where, in particular, 
 the {\bf limiting amplitude principle} (\ref{lap}) was established as well as (\ref{lab}).

Another approach to the construction of solutions to (\ref{AB}) has been
suggested by   Maz'ya and Plamenevskii \ci{MP1971, MP1975}.
This approach is applicable only to equations with real coefficients
that is not sufficient for application to the diffraction problems.

Many works   published since 1980' concern 
a wide spectrum of
 properties of solutions to the boundary problems of type
(\ref{AB}) in different regions with angles, see Grisvard \ci{G1985}, Costabel and  
Stephan \ci{CS1985}, Dauge \ci{D1988}, Bernard \ci{B1987,B2006}, Nazarov and Plamenevskii \ci{NP1994}, Bonnet-Ben Dhia and Joly \ci{BJ1993}, 
Bonnet-Ben Dhia, Dauge and Ramdani \ci{BDR1999}, Meister with collaborators \ci{M1987}--\ci{M1992}, 
Penzel  and Teixeira \ci{PT1999}, Castro and Kapanadze \ci{CK2015}, and others. The detailed
survey can be found in \ci{KM2019}.

  Note that Malyshev's method plays an important role in the theory
of Queueing Systems and Analytic Combinatorics \ci{FIM2017}.

Another  important area of application of Malyshev's method 
is the linear theory of water waves. In particular,
the method was applied in 1996--2002  by the authors together with P.N. Zhevandrov
to trapped modes on a sloping beach. 
As the result, the long-standing problem of the completeness of the Ursell's modes 
has been solved
\ci {KMZ1996,M1995}, (see also \ci {MZ1998, ZM} where this method was  used ). This progress is due to the fact
that the method allows one to obtain {\it all} solutions of the boundary value 
problems in angles.

We expect  that  the method can give a valuable progress in diffraction
by ferromagnetic wedges which is a challenging open problem
of  radar  detection. In this case, the operators $B_l$ in  (\ref{AB}) 
are nonlocal pseudodifferential operators.


\section{Malyshev's method of automorphic functions}

In this section, we present basic steps of the method \ci{K1973}
which relies on Malyshev's ideas of automorphic functions \ci{M1970}.

Note that in the case of rational angles $\Phi=\pi/n$ and
the Dirichlet and Neumann boundary conditions,
the boundary value problem (\ref{Helm})
 can be  easily solved by  reflections in the sides
of the angle. This method was well known  at least 
since the Gauss theory of electrostatics \ci{G1839}.
For  $\Phi\ne \pi/n$ the reflections do not give a solution, and
for  irrational $\Phi/\pi$, the method suggested the reflections on 
a ``Riemann surface" formed by the reflected angles. This was the original 
step of the Sommerfeld approach which leaded him to the famous
``Sommerfeld integral representation" for solutions \ci{S1896}. 
The reflection
on the Riemann surface and the theory of branching solutions to the wave equation
have been developed later by Sobolev \ci{S1935} 
and \ci[Chapter XII]{S1937}.

Very   surprisingly, the  method of automorphic functions \ci{K1973, M1970} 
also 
 relies on the reflections on a suitable Riemann surface $V$. However, in this approach, 
 $V$ is the   surface in the Fourier space, contrary to the original ideas
 of Sommerfeld. Namely, $V$ is the 
Riemann surface of complex
characteristics of the elliptic operator $A$:
\be\la{V}
V=\{z\in\C^2: \hat A(z)=0\}.
\ee
\br
\rm
The main idea of the Malyshev  approach is the invariance of the 
Cauchy data of solutions under covering maps of the Riemann surface $V$,
see Remark \ref{rMal}.
\er

In \ci{K1973}, the problem (\ref{AB}) 
 with strongly--elliptic operators $A$ in {\bf convex angles} $Q$
 is solved in the following steps:
\smallskip\\
1. Reduction to an undetermined algebraic equation
with two unknown functions
 on the Riemann surface $V$.
2. Elimination of one unknown function using its   invariance with respect to 
the covering map of the Riemann surface.
\smallskip\\
3.  Reduction of the obtained equation with one unknown 
function to the Riemann--Hilbert problem on $V$.
\smallskip

Below in this section, we describe some details.
 
\subsection{Reduction to undetermined algebraic equation on the Riemann surface}

As an example, we consider the Dirichlet boundary value problem
in the quadrant $Q=\R^+\times\R^+$ :
\be\la{AB2}
\left\{\ba{rcl}
A u(x_1,x_2)&=&0\\\\
 u(x_1,0)&=&f_1(x_1),\,\,  u(0,x_2)=f_2(x_2)
  \ea\right|,\,\,\quad x_1>0,\,\,\,x_2>0.
    \ee
    
\subsubsection{Fourier--Laplace transform}
We assume that the solution $u(x)\in C^2(\ov Q)$ and 
is bounded by a polynomial:
\be\la{upol}
|u(x)|+|\na u(x)|\le C(1+|x|)^p,\qquad x\in\R^2.
    \ee
    Denote $\C^+=\{\zeta\in\C:\rIm \zeta>0\}$ and 
    $Z^+=\C^+\times\C^+$, and
    consider
the complex  Fourier--Laplace transform of solution
 \be\la{FL}
 \hat u(z)=\int_0^\infty \int_0^\infty  e^{izx}u(x)dx_1dx_2,\qquad z=(z_1,z_2)
 \in Z^+.
 \ee   
 By (\ref{upol}),
 this integral is absolutely   convergent  and hence  it
is an analytic function of two complex variables
(this is a particular case of the Paley--Wiener Theorem).
Let us denote the Neumannn 
data of the solution as
 \be\la{Nd}
 \vp_1(x_1)=\pa_2u(x_1,0),\,\,\,x_1\ge 0;\qquad
 \vp_2(x_2)=\pa_1u(0,x_2),\,\,\,x_2\ge 0.
\ee
It is well known that the solution $u(x)$ can be expressed  via the Dirichlet and Neumannn data
$f_1,f_2,\vp_1,\vp_2$ by the    Green integral formula \ci{CH}. 
In our case, it is useful to obtain this formula in the Fourier transform.
For this purpose, multiply the first equation in (\ref{AB})
by $e^{izx}$    and integrate over $Q$. Integrating by parts, we immediately obtain
\be\la{4}
0=\int_0^\infty\int_0^\infty e^{izx}Au(x)dx_1dx_2
=
\hat A(z)\hat u(z)+F(z),\qquad z\in Z^+,
\ee
 where 
 \be\la{F}
 F(z)=P_1(z)\hat f_1(z_1)+P_2(z)\hat f_2(z_2)+S_1(z)\hat \vp_1(z_1)+S_2(z)\hat \vp_2(z_2),\qquad z\in Z^+,
 \ee
and the functions $P_l$ and $S_l$ are  polynomials.

\subsubsection{Undetermined algebraic equation on the Riemann surface}

Rewrite (\ref{4}) as
\be\la{42}
\hat A(z)\hat u(z)=-F(z),\qquad z\in Z^+.
\ee
Now (\ref{V}) implies the identity
\be\la{FVK}
F(z)=0,\qquad z\in V^+:=V\cap Z^+
\ee
since all the functions $\hat A(z),\hat u(z),F(z)$  are {\bf 
analytic} in the domain $Z^+$! 
\br
{\rm
Note that the set of complex characteristics 
$V$ is nonempty even for strongly elliptic operators (\ref{sel}),
though its intersection with the real plane $\R^2$ is empty; see Example
\ref{exem} below. 
}
\er
The identity (\ref{FVK}) can be rewritten as {\bf undetermined linear algebraic equation}
\be\la{aleq7}
 S_1(z)\vp_1(z_1)+S_2(z)\vp_2(z_2)=G(z),\qquad z\in V^+
\ee
with {\it\bf two unknown functions} $\vp_1(z_1)$, $\vp_2(z_2)$,
and with known right-hand side:
\be\la{G}
G(z):=-P_1(z)\hat f_1(z_1)-P_2(z)\hat f_2(z_2),\qquad z\in V^+.
\ee

\br{\rm
The identity 
(\ref{42}) implies the formula for the solution
\be\la{43}
 u(x)=-\Big[\cF^{-1}\fr{F(z)}{\hat A(z)}\Big](x),\qquad x\in Q,
\ee
where 
$\cF^{-1}$ denotes the inverse to the Fourier--Laplace transform
(\ref{FL}), and
the right hand side is well defined due to (\ref{sel}).
The formula (\ref{43}) can be transformed into  the well known Green formula
which expresses the solution $u(x)$  via its Cauchy data. 
}\er

\subsection{Method of automorphic functions}\la{saf}

\subsubsection{Covering maps}
Denote 
\be\la{psifg}
\psi_1(z)=\vp_1(z_1), \quad\psi_2(z)=\vp_2(z_2),\qquad
\hat g_1(z)=\hat f_1(z_1),\quad\hat g_2(z)=\hat f_2(z_2).
\ee
  Now
 (\ref{aleq7}) becomes
\be\la{aleq9}
 S_1(z)\psi_1(z)+S_2(z)\psi_2(z)=G(z),\qquad z\in V^+,
\ee
where 
\be\la{G2}
G(z):=-P_1(z)\hat g_1(z)-P_2(z)\hat g_2(z).
\ee
Of course, this equation is not equivalent to  (\ref{aleq7}). 
To keep the equivalence, 
we need
an additional characterisation of the functions $\psi_l(z)$.
This is the key  observation of Malishev that the functions are automorphic 
with respect to  an appropriate groups of transformation of the Riemann surface $V$.

\begin{figure}[htbp]
		\begin{center}
			\includegraphics[width=0.68\columnwidth]{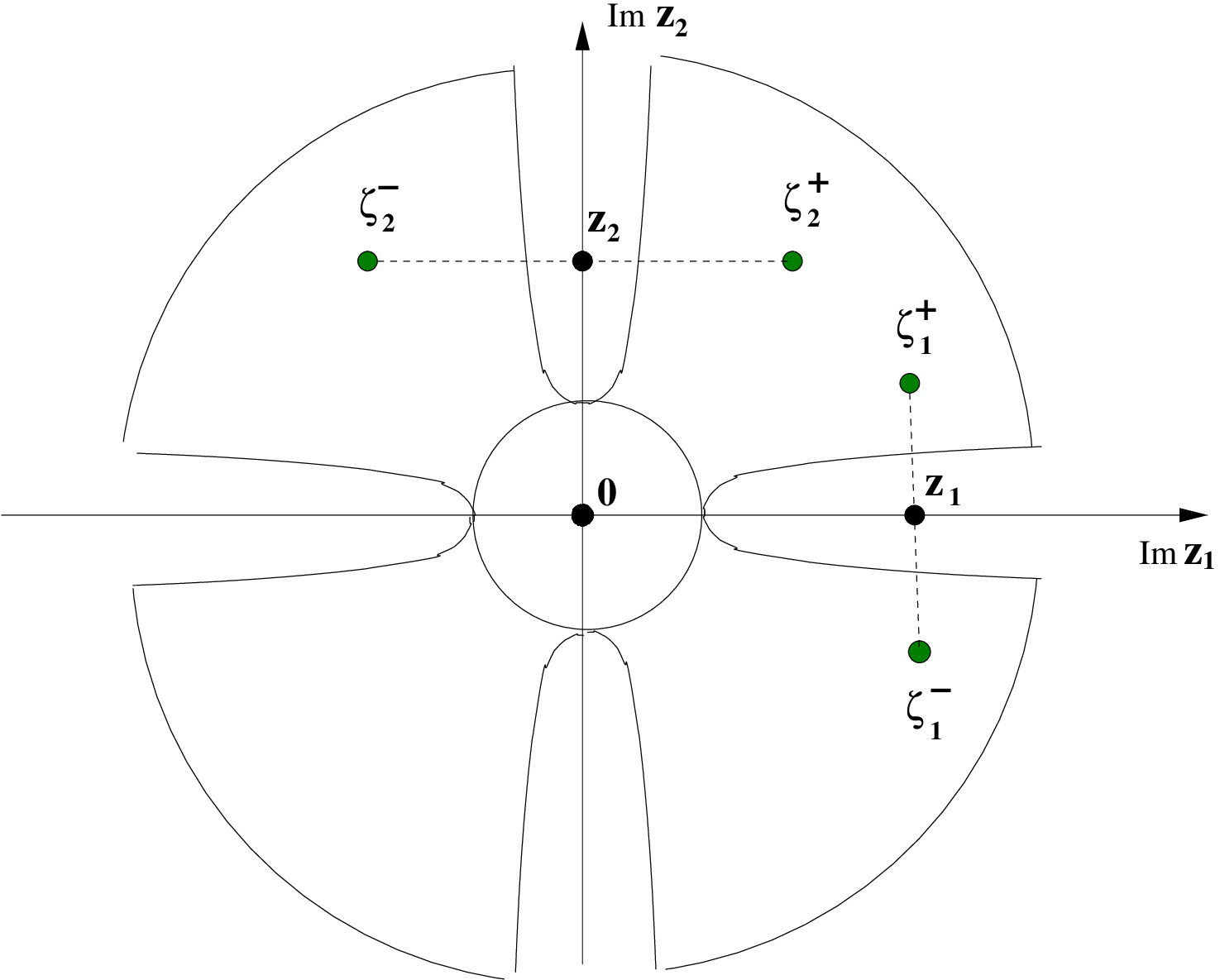}
			\caption{Riemann surface $V\!\!: z_1^2+z_2^2+1=0$
			in projection onto the plane $\rIm z_1,\rIm z_2$.
			}
			\label{fig3}
		\end{center}
	\end{figure}

First, consider the
coordinate projections $p_l:V\to \C$ defined by
\be\la{copr}
p_1(z_1,z_2)=z_1, \qquad p_2(z_1,z_2)=z_2.
\ee
These projections 
 are two-sheeted since, for example, $p_1(z_1,z_2)=z_1$ means that
 $z_2$ is the root of the quadratic equation $\hat A(z_1,z_2)=0$.
Accordingly, the inverse maps $p_l^{-1}:\C\to V$ are  double-valued:
for $z_1,z_2\in\C$,
\be\la{inver}
p_1^{-1}(z_1)=\{\zeta_1^-,\zeta_1^+\},\qquad
p_2^{-1}(z_2)=\{\zeta_2^-,\zeta_2^+\},
\ee
and at the branching points of $p_l^{-1}$, the two points $\zeta_l^\pm\in V$  coincide.
\bd
{\bf\em  Covering maps} $h_1,h_2:V\to V$ are defined as follows: for any $z_1,z_2\in\C$,
\be\la{covm}
h_1\zeta_1^\pm=\zeta_1^\mp,  \qquad\qquad h_2\zeta_2^\pm=\zeta_2^\mp.
\ee
\ed
\bex\la{exem}
{\rm
For the strongly-elliptic operator $A=-\De+1$,
the corresponding Riemann surface $V\!\!: z_1^2+z_2^2+1=0$
 is  shown in Fig. \ref{fig3} in projection onto the plane $\rIm z_1,\rIm z_2$. It is easy to see that this projection does not cover the circle $|\rIm z_1|^2+|\rIm z_2|^2< 1$, and it covers twice each point
with $|\rIm z_1|^2+|\rIm z_2|^2>1$.
The surface
consists of two sheets shown  in  Fig. \ref{fig3}, and glued along the cuts.

}
\eex

		Thus, $h_1$ permutes the points $\zeta_1^\pm\in V$ with 
 the  identical projections $z_1=p_1\zeta_1^\pm$, and similarly,  $h_2$ permutes the points $\zeta_2^\pm\in V$ with  the  identical projections $z_2=p_2\zeta_2^\pm$ (see Fig. \ref{fig3}):
\be\la{ph}
p_1h_1\zeta_1^\pm=p_1\zeta_1^\mp=z_1,\qquad
 p_2h_2\zeta_2^\pm=p_2\zeta_2^\mp=z_2.
\ee
The maps $h_l:V\to V$ with $l=1,2$ define the corresponding 
 automorphisms of the ring of (meromorphic) functions $\psi(z)$ on the Riemann
 surface $V$:
\be\la{autV}
\psi^{h_l}(z):=\psi(h_lz), \qquad z\in V.
\ee

{ Figure \ref{fig3} shows that
$$
\ba{ll}
p_1\zeta_1^+=z_1=p_1\zeta_1^-,& {\rm so}\,\,\,
			\psi_1(\zeta_1^+)= \vp_1(z_1)=\psi_1(\zeta_1^-),
			\\\\
			 p_2\zeta_2^+=z_2=p_2\zeta_2^-,& {\rm so}\,\,\,
			\psi_2(\zeta_2^+)= \vp_2(z_2)=\psi_2(\zeta_2^-).
\ea
$$
	\bigskip
Now it is clear that 
the functions $\psi_l(z):=\vp_l(z_l)$ with $l=1,2$ are invariant with respect to 
the automorphisms $h_l$:
}
\be\la{aut}
\,\,\,\,\,
\psi_l^{h_l}(z)=\psi_l(z), \qquad\,\,\,\,\,\,\,z\in V^+.
\ee
In other words,  the functions $\psi_l$ are {\bf automorphic}, and
the automorphisms defined by
$h_l$ belong to the corresponding {\bf Galois groups} of extensions
of the ring of functions of $z_l$.

\subsubsection{ Shift equation}
 
Applying {\bf formally} $h_1$ to (\ref{aleq9}), and using (\ref{aut}) with $l=1$, we get {\bf   a new equation} for the same  unknown functions:
\be\la{aleq10}
S_1^{h_1}(z)\psi_1(z)+S_2^{h_1}(z)\psi_2^{h_1}(z)=G^{h_1}(z).
\ee
The problem is that 
$\psi_2^{h_1}(z)=\psi_2(h_1z)$
and
$G^{h_1}(z)=G(h_1 z)$   
 are not defined generally  for $z\in V^+$ since $V^+$ is not invariant with respect
to the covering map $h_1$. In particular, we have by (\ref{G2}),
\be\la{Gh1}
G^{h_1}(z):=-P_1^{h_1}(z)\hat g_1(z)-P_2^{h_1}(z)\hat g_2^{h_1}(z),
\ee
where $\hat g_2^{h_1}(z)=\hat g_2(h_1 z)$   is not defined generally for $z\in V^+$.
To save the situation, consider the case $f_2=0$. Then $\hat g_2=0$,  and now
the right hand side of equation (\ref{aleq10}) is well defined for $z\in V^+$.
It is important that in this case $\psi_2^{h_1}(z)$ is also well defined 
\ci[Ch. 14]{KM2019}.
The case $f_1=0$
can be considered similarly.

\br
{\rm
The function $\psi_2(z)$ admits an analytic continuation outside 
the region $V^+_l:=\{z\in V:\rIm z_2>0\}$ {\it on the Riemann surface} $V$, see \ci[Ch. 14]{KM2019}. 
Let us stress that
this is analytic continuation {\it along the surface} $V$.
}
\er

Now we can {\bf eliminate} the function $\psi_1$ from (\ref{aleq9}) and  (\ref{aleq10}). As a result, we obtain
 an algebraic equation with a shift for {\bf one unknown function}
\be\la{aleq3}
R_1(z)\psi_2^{h_1}(z)-R_2(z)\psi_2(z)=H(z),\quad z\in V^+.
 \ee
 Finally, using (\ref{aut})  {\bf\em with l=2}, we get 
\be\la{aleq32}
R_1(z)\psi_2^{h}(z)-R_2(z)\psi_2(z)=H(z),\quad z\in V^+;\qquad h=h_2h_1.
 \ee

 \br\la{rMal}
 \rm
 The elimination of unknown functions using their invariance with respect
 to suitable ``reflections" is the main idea of Malyshev's method.
 
 \er
 
 \subsection{Reduction to the Riemann--Hilbert problem}\la{srh}
 Let us illustrate the reduction of  equation (\ref{aleq3})
 to the Riemann--Hilbert problem
for a particular case of strongly-ellipltic operator $A=-\De+1$. Its symbol is 
 $\hat A(z)=z^2+1$, so
$V=\{(z_1,z_2)\in \C^2: z_1^2+z_2^2=-1\}$ and the covering maps are
\be\la{Ade}
h_1 (z_1,z_2)=(z_1,-z_2),\qquad h_2 (z_1,z_2)=(-z_1,z_2).
\ee
 Introduce the coordinate $w$ on the universal covering $\hat V=\C$
 of the surface $V$ by 
 \be\la{theta}
 z_1=i\cos w,\qquad z_2=i\sin w.
 \ee
 The maps (\ref{Ade}) can be lifted to $\hat V$
 as
 \be\la{Adew}
\hat h_1 w=-w,\qquad \hat h_2 w=-w+\pi .
\ee
  Now $h=w+\pi$, so (\ref{aleq32}) becomes
 \be\la{aleq33}
\ti R_1(w)\ti \psi_2(w+\pi)-\ti R_2(w)\ti \psi_2(w)=\ti H(w).
 \ee
 where $\ti R_1$, etc,  denote the liftings of the corresponding
 functions  to the universal covering. The equation (\ref{aleq33}) holds
 for an appropriate region of $w\in\C$. Restricted to the strip $\rRe w\in [0,\pi]$,
 this equation is the Riemann--Hilbert problem which can be solved in quadratures
 \ci[Chs 17 and 18]{KM2019}. Let us recall some details.

The function $z=e^{2i w}$ analytically 
 transforms  the strip to the plane with the cut $[0,\infty)$. 
 Denote the function $\check\psi_2(t)=\ti\psi_2(w)$,
 $\check H(t)=\ti H(w)$ and 
 $\check R_k(t)=\ti R_k(w)$ for $k=1,2$. 
 Then  relation (\ref{aleq33})
 becomes
  \be\la{aleq332}
\check R_1(t)\check \psi_2(t-i0)-\check R_2(t)\check \psi_2(t+i0)=\check H(t),
\qquad t>0.
 \ee
 As the first step of the  Riemann--Hilbert method, one must solve the 
 corresponding homogeneous problem:
  \be\la{aleq332T}
\check R_1(t) T(t-i0)-\check R_2(t)\check T(t+i0)=0,
\qquad t>0.
 \ee
 Equivalently, 
 \be\la{aleq332Te}
 \fr{T(t+i0)}{T(t-i0)}=q(t):=\fr{\check R_1(t) }{\check R_2(t) },
\qquad t>0.
 \ee
 The solution to this equation depends on zeros of the functions 
 $\check R_1(t)$ and $\check R_2(t)$  for $t>0$. 
 Let us consider the simplest case when such zeros do not exist, and moreover,
\be\la{qq} 
 q(0)=q(\infty)=1.
 \ee
 Then
 the equation is equivalent to
 \be\la{aleq332T2}
\log T(t+i0)-\log T(t-i0)=\log q(t),
\qquad t>0.
 \ee
 The solution
 is given by the Cauchy-type integral 
 \be\la{aleq332T3}
\log T(t)=\fr1{2\pi i}\int_0^\infty \fr{\log q(s)}
{t-s} ds,
\qquad t\in \C\setminus[0,\infty).
 \ee
 It is important that $T(t)$ is analytic and nonvanishing in the region 
 $ \C\setminus[0,\infty)$.
 Now the  nonhomogeneous problem (\ref{aleq332}) can be solved as follows.
 First,  (\ref{aleq332}) and (\ref{aleq332T}) imply
  \be\la{aleq332i}
\fr{\check \psi_2(t-i0)}{T(t-i0)}-
\fr{\check \psi_2(t+i0)}{T(t+i0)}
=\fr{\check H(t)}{\check R_1(t) T(t-i0)},
\qquad t>0.
 \ee
 Therefore, similarly to (\ref{aleq332T2}),
 \be\la{aleq332is}
\fr{\check \psi_2(t)}{T(t)}
=
-
\fr1{2\pi i}\int_0^\infty
\fr{\check H(s)}{\check R_1(s) T(s-i0)(t-s)}ds,
\qquad t\in \C\setminus[0,\infty)
 \ee
 since the function $\fr{\check \psi_2(t)}{T(t)}$ is analytic in 
 $\C\setminus[0,\infty)$.

 Thus, we have calculated the function  $\check \psi_2(t)$.
 Now $\psi_2(z)$ can be obtained from the relation (\ref{aleq10}). 
Hence, the functions  $\hat\vp_1(z_1)$ and $\hat\vp_2(z_2)$ are known.
 It remains to
 substitute
 the obtained functions
 into  the formula (\ref{F}) for the function $F$. Then the solution to (\ref{AB}) 
  is expressed by (\ref{43}), which can be reduced to the integral of Sommerfeld type
  \ci{KMM2005}.
 
 \br
 {\rm
   Equation (\ref{aleq3}) is obtained using the invariance 
  (\ref{aut}) with $l=1$, while (\ref{aleq32}) uses also $l=2$.
  Note that
 the equation (\ref{aleq3})  reads now 
  \be\la{aleq34}
\ti T_1(w)\ti \psi_2(-w)-\ti T_2(w)\ti \psi_2(w)=\ti H(w),
 \ee
 which provisionally 
 cannot be reduced to a nonsingular Riemann--Hilbert problem,
see \ci{L1977}. Thus, both invariance conditions (\ref{aut}) are necessary
for the reduction.
 
 }
 
 \er

   \br
  {\rm
  For the random walks studied  in 
  \ci{M1970, FIM2017},
  the corresponding 
  Riemann surface and 
  the  covering maps $h_l$ can be more complicated than for 2-nd order elliptic operators  which requires more sophisticated methods of the Galois theory.
  }
  \er

 \section{Nonconvex angles of magnitude $\Phi>\pi$}
 
 The extension of the theory  outlined  above to the
 case of nonconvex angle $Q$
  differs drastically from the convex one. 
 As an example, consider the Dirichlet boundary value problem
in the angle $Q=\R^2\setminus \R^+\times\R^+$ :
\be\la{AB3}
\left\{\ba{rcll}
A u(x)&=&0, \qquad x\in Q&
\\\\
 u(x_1,0)&=&f_1(x_1), \,\,x_1>0;& u(0,x_2)=f_2(x_2),\,\,x_2>0.
  \ea\right|.
    \ee
Note that
the relations (\ref{4}) and (\ref{42}), (\ref{43}) 
 remain true in this case, but now the function (\ref{F})  is changed to  
 its negative:
 \be\la{F2}
 F(z)=-P_1(z)\hat f_1(z_1)-P_2(z)\hat f_2(z_2)-S_1(z)\hat \vp_1(z_1)-S_2(z)\hat \vp_2(z_2),\qquad z\in\R^2.
 \ee
 { On the other hand, the key relation (\ref{FVK}) is not well defined
 in contrast to the case when the support of $u$ belongs to  a convex angle.
 This is due to the fact  that 
 the Fourier--Laplace transform (\ref{FL}) of the function $u$ with the support in a nonconvex angle
 generally does not admit  an analytical continuation to a region of  $\C^2$.
 Nevertheless, the function (\ref{F2})  in this case  
  is analytic in the same region 
$Z^+=\C^+\times\C^+$ as the function (\ref{F}).
}

 The answer to this riddle was found in \ci{KM1992}. 
First,  the function (\ref{F2}) admits the splitting
 \be\la{split}
 F(z)=\ga_1(z)+\ga_2(z), \quad \ga_1(z)=-P_1(z)\hat f_1(z_1)-S_1(z)\hat \vp_1(z_1),\quad \ga_2(z)=-P_2(z)\hat f_2(z_2)-S_2(z)\hat \vp_2(z_2),
 \ee
 where the functions $\ga_l(z)$ are analytic in the regions 
 \be\la{Vpm} 
 V_l^+=\{z\in V: \rIm z_l>0\},\qquad l=1,2.
 \ee
Second, as   shown in \ci{KM1992} (see also \ci[Theorem 20.1]{KM2019}), 
each function $\ga_l$ admits an analytic continuation 
from $V_l^+$ to the region $V^-:=\{z\in V:\rIm z_1<0,\,\rIm z_2<0 \}$, and
the following identity holds:
\be\la{}
\ga_1(z)+\ga_2(z)=0,\qquad z\in V^-.
\ee
This identity
 formally coincides with the undetermined equation (\ref{aleq7}),
 and it
 allows  us to calculate both unknown functions $\hat\vp_l$
by methods of Sections \ref{saf} and \ref{srh}.

 \section{Time-dependent diffraction by wedge}
 
 The time-dependent diffraction by a wedge $W$ is described by  the solution
 of the wave equation in the plane angle $Q=\R^2\setminus W$ with appropriate boundary conditions.
 For example, consider the Dirichlet boundary conditions  
 \be\la{nsd}
 \left\{\ba{rcl}
 \ddot u(x,t)&=&\De u(x,t) ,\quad x\in Q
 \\\\
 u(x,t)&=&0, \quad x\in\Ga_1\cup\Ga_2
 \ea\right|,  \quad t\in\R.
 \ee
 The incident wave is defined by the initial condition
 \be\la{ini}
 u(x,t)=u^{in}(x,t), \qquad u^{in}(x,t):=f(kx-\om_0 t)e^{i(kx-\om_0 t)},\qquad t<0,
 \ee
 where 
  the frequency $\om_0\in\R$
 and
 $k\in\R^2$ is the {\bf wave vector}.
 The incident wave $u^{in}(x,t)$
 must be a solution to (\ref{nsd}) for $t<0$:
 \be\la{nsd1}
 \left\{\ba{rcl}
 \ddot u^{in}(x,t)&=&\De u^{in}(x,t) ,\quad x\in Q
 \\\\
 u^{in}(x,t)&=&0, \quad x\in\Ga_1\cup\Ga_2
 \ea\right|,  \quad t<0.
 \ee
The wave equation in  (\ref{nsd1}) holds
for any function $f(s)$
{\bf for all $t\in\R$}
 if $|k|=|\om_0|$.
The boundary condition in  (\ref{nsd1}) can be satisfied 
only in the case of nonconvex angle $Q$ of   magnitude $\Phi>\pi$
and the 
 wave vector $k$  satisfying the inequalities 
 $k\cdot x\ge 0$ for $x\in W=\R^2\setminus Q$.
Then for $\om_0>0$
 the boundary condition holds if
\be\la{as}
f(s)=0,\qquad s>0.
\ee

\section{Limiting amplitude principle}\la{std}

Let us  assume that there exists the limit
\be\la{alim}
 f(-\infty):=\lim_{s\to-\infty}f(s),
 \ee
 and the convergence is sufficiently 
fast, 
 for example, $f(s)=\theta(-s)$.
 Then the incident wave (\ref{ini})
 admits the long-time asymptotics
 \be\la{ini2}
 u^{in}(x,t)\sim f(-\infty)e^{ikx} e^{-i\om_0 t},\qquad t\to\infty
 \ee
 which suggests similar  asymptotics
of solution
 \be\la{lap}
u(x,t)\sim a_{\om_0}(x)e^{-i\om_0 t},\qquad t\to\infty.
 \ee
 Such asymptotics are 
 called as {\bf limiting amplitude principle}.

 Determination of the {\bf limiting amplitudes} $a_{\om_0}(x)$ 
 for different  diffraction processes
 is the main goal of the theory of diffraction \ci{BW1966,S1954}  (see also \ci{KM2019}).
 The proof of the asymptotics is the main goal of the  mathematical
 theory of diffraction. For diffraction by wedges, this asymptotics
 has been established for the first time in \ci{KM2006}.
 Formal substitution of the asymptotics (\ref{lap}) into (\ref{nsd}) gives
 a problem of type (\ref{Helm}):
 \be\la{nsd2}
 \left\{\ba{rcl}
 -\om_0^2a_{\om_0}(x)&=&\De a_{\om_0}(x) ,\quad x\in Q
 \\\\
 a_{\om_0}(x)&=&0, \quad x\in\Ga_1\cup\Ga_2
 \ea\right|.
 \ee
  However,
 this boundary problem is {\bf  ill-posed} 
 since it admits an infinite number of linearly independent 
solutions
 for real $\om_0\in\R$.
 Thus, this problem 
 does not allow   us to find the limiting amplitude.
 This fact is the main  peculiarity of the diffraction theory.
 This  can be easily checked  in the case $\Phi=\pi$ when the angle
$Q$ is the half-plane, so all solutions can be calculated by the Fourier transform
along the boundary $\pa\Om$. For the problems of type (\ref{nsd2})
in {\bf convex angles} of magnitude $\Phi<\pi$, this nonuniqueness was discovered in 1973 by one of the authors \ci{M1977}.
 \smallskip

 Let us recall how to prove the asymptotics (\ref{lap})
  and  how to calculate the limiting amplitudes $a_{\omega_0}(x)$.
  First, 
 note that for the incident wave $u^{in}(x,t)$
 the asymptotics of type (\ref{lap}) holds  
 by (\ref{alim}):
 \be\la{lapin}
u^{in}(x,t)\sim f(-\infty)e^{ikx}e^{-i\om_0 t},\qquad t\to\infty.
 \ee 
 The reflected wave is defined 
 by geometric optics, and its main 
 properties are as follows:
 \be\la{refl} 
  u^r(x,t)=-u^{in}(x,t),\quad x\in\pa Q;\qquad
 u^{r}(x,t)\sim a^r(x)e^{-i\om_0 t},\qquad t\to\infty.
 \ee
 The diffracted wave $u^d(x,t)$ is defined by the splitting
  the  total solution as
 \be\la{spli}
 u(x,t)=u^{in}(x,t)+u^r(x,t)+u^d(x,t).
 \ee
 Hence, it remains
  to calculate the corresponding  asymptotics for the diffracted wave
 \be\la{laps}
u^d(x,t)\sim a^d_{\om_0}(x)e^{-i\om_0 t},\qquad t\to\infty,
 \ee
 Substituting (\ref{spli}) into (\ref{nsd}), 
 using  (\ref{refl}) and the fact that the wave equation in (\ref{nsd1}) holds for all $t\in\R$,
 we get
  the boundary problem for the diffracted wave
 \be\la{nsd3}
\left\{\ba{rcl}
 \ddot u^d(x,t)&=&\De u^d(x,t) +F(x,t),\quad x\in Q
 \\\\
 u^d(x,t)&=&0, \quad x\in\Ga_1\cup\Ga_2
 \ea\right|,\qquad F(x,t):=(\pa_t^2-\De)u^r(x,t)\sim b(x)e^{-i\om_0t},\,\,\,t\to\infty.
 \ee 
 Formal substitution of the asymptotics (\ref{laps}) into
 (\ref{nsd3}), gives the boundary problem
 \be\la{nsd4}
\left\{\ba{rcl}
 -\om_0^2 a^d_{\om_0}(x,t)&=&\De a^d_{\om_0}(x) +b(x),\quad x\in Q
 \\\\
 a^d(x)&=&0, \quad x\in\Ga_1\cup\Ga_2
 \ea\right|.
 \ee 
 For $\om_0\in\R$, this system also admits an infinite number of linearly independent solutions,
 as well as (\ref{nsd2}). Similar problem of nonuniqueness arises in every
 diffraction problem in unbounded regions. 
 The problem of nonuniqeness was resolved 
 by  the discovery of additional 
 features of the limiting amplitude $a^d_{\om_0}(x)$.
  
 The key discovery was the {\bf limiting absorption principle} (\ref{lab}) for the  limiting amplitude of the diffracted wave.
  In application to problem  (\ref{nsd4}),  we have
  \be\la{labs}
 a^d_{\om_0}(x)=\lim_{\ve\to 0+}a^d_{\om_0+i\ve}(x),\qquad x\in Q,
\ee
 where $a^d_{\om_0+i\ve}$ denotes a solution to (\ref{nsd4}) with 
$\om_0+i\ve$ instead of $\om_0$.

\br\rm
The convergence (\ref{labs}) holds for the limiting amplitude
$a^d_{\om_0}(x)$ of the  diffracted wave $u^d_{\om_0}(x,t)$
(formal proof can be found in \ci[Section 4.1]{KM2019}).
However, 
it does not hold for the  limiting amplitude $a(x)$
of the total solution $u(x,t)$
 although
these amplitudes satisfy  quite similar equations
(\ref{nsd4}) and (\ref{nsd2}). The difference is
that the initial state of the diffracted wave $u^d_{\om_0}(x,0)$ is 
 of finite energy (in our case zero), while for the total solution
 the initial state $(u(x,0),\dot u(x,0))$ is the plane wave (\ref{ini}) 
and its derivative in time 
at $t=0$.
\er

 The  limiting absorption principle has been introduced  for the first time   in 1905 by W. Ignatovsky \ci{I1905}.  Rigorous proofs of this principle 
 for limiting amplitudes of solutions with 
 finite energy initial states
 were achieved much later.
The results for the wave and Schr\"odinger equations
 in the entire space and for diffraction problems
 with
 smooth boundaries were obtained by Agmon \ci{A1975}, 
 Eidus \ci{E1965,E1969,E1989}, Jensen and Kato \ci{JK1979},  A.Ya. Povzner \ci{P1953},  B.R. Vainberg \ci{V1966} and others.

 The convergence (\ref{labs}) for {\bf stationary diffraction problems}
has been established for the  first time in 1977 by one of the authors 
 \ci{M1977}: 
  it was proven that
stationary problem    (\ref{nsd4}) and 
problems (\ref{AB}) with $A=\De+\om^2$ and 
 general boundary conditions in {\bf convex angles} of magnitude $\Phi<\pi$,
\smallskip\\
i) for complex $\om\not\in\R$ admit only a 
finite number of linearly independent solutions in appropriate 
class of functions; 
\smallskip\\
ii) for real $\om\in\R$
admit an infinite number of linearly independent solutions,
\smallskip\\
iii) for real $\om\in\R$
 admit only a {\bf finite number} of linearly independent solutions satisfying (\ref{labs}).
\smallskip

 For the {\bf time-dependent} diffraction problem (\ref{nsd}), (\ref{ini}),
 the limiting absorption principle (\ref{labs}) and 
 the limiting amplitude principle (\ref{lap})
  were justified in 2006  by the authors \ci{KM2006}. 
 The proofs rely on the  analysis of
 the Fourier-Laplace transform in time:
 \be\la{Ft}
 \ti u(x,\om)=\int_0^\infty e^{i\om t} u(x,t)dt,\qquad \om\in\C^+.
 \ee
 The function $\ti u(x,\om)$ satisfies  a boundary value problem
 of type
 (\ref{nsd4}) with complex $\om\not\in\R$. 
 In this case the Helmholtz operator $A=\De+\om^2$
 is  strongly elliptic. Hence,
 $\ti u(x,\om)$ can be
   calculated and analysed by the methods described in previous sections.
   The limiting amplitude is calculated in  \ci{KM2006} using 
   the limit (\ref{labs}).
   
 \br 
 \rm
  In 1912,  A. Sommerfeld  discovered the {\bf Sommerfeld radiation condition}
 \ci{S1912}
(see also \ci{S1992}),
 which  provides the uniqueness of solution to the boundary problem
  of type  (\ref{nsd4}) 
 in the case
 when $Q$ is the exterior of a bounded region in $\R^3$.
 This condition is more practical for numerical calculation
 of the limiting amplitudes than (\ref{labs}).

 \er

 \section{The Sommerfeld diffraction theory and related results}

 For the angle $\Phi=2\pi$,
  A. Sommerfeld constructed in 1896 a solution $a(x)$ to {\bf stationary
 diffraction problem} of type (\ref{nsd2}) 
 with Dirichlet and Neumannn boundary conditions. In this case the wedge is the half-plane, 
 which is represented by the semi-axis  $[0,\infty)$ in the corresponding 
 2D problem.
 The main ideas were i) to treat the  semi-axis as the cut on an appropriate Riemann
 surface, and ii) to extend the known method of reflections to Riemann
 surfaces.
 As a result, A. Sommerfeld constructed a universal integral representation
of a class of {\bf branching solutions} of the Helmholtz equation 
on the Riemann surface
in the form of the 
{\bf Sommerfeld integral} with a fixed integral kernel 
and a with a suitable density function. 
Further, \linebreak A. Sommerfeld  chose an appropriate
 densities to satisfy the boundary conditions.
 
 Sommerfeld's strategy  of  constructing the solution
remains a mysterious riddle to this day.
This approach is reproduced with some comments in \ci[Ch. 5]{KM2019}, see also \ci{NZS2004}. 
However, the Sommerfeld integral representation
  turned out to be  extremely  fruitful, and in particular, was used by G.D. Malujinetz
  to solve the problem with the Leontovich boundary condition \ci{M1958,M1959}, see also \ci{BLG2007}.
  
  For any angles $\Phi\in (0,2\pi)$
    the {\bf stationary diffraction problem} (\ref{nsd2}) 
  for the Dirichlet and Neumannn boundary conditions
  in the angles of  this  magnitude 
  was solved by  other methods in 1920 by  H.S. Carslaw \ci{C1920},
  in 1932--1937 by
 V.I. Smirnov and  S.L. Sobolev \ci{SS1932,S1934,S1935,S1937,S1958}, and in 1951 by J.B. Keller and  A. Blank \ci{KB1951}.

 \br
 {\rm

i)  In all  the works, cited above, the limiting amplitude principle was not established, and 
the choice of suitable solution of the {\bf  ill-posed problem} (\ref{nsd2}) was not rigorously clarified.
  Nevertheless, 
   as shown in \ci{KM2015,KMNMV2018,MKMV2015},
  all  the obtained solutions coincide with the {\bf limiting amplitudes}
  calculated in \ci{KM2006}
  and admit the Sommerfeld representation.
  \smallskip\\
ii) S.L. Sobolev mentions, in the articles cited above,
that the functions of type (\ref{ini})
must be solutions to the wave  equation even if the amplitude $a(s)$ is a discontinuouos function. These remarks later inspired  the theory of weak derivatives of S.L. Sobolev
and the theory of distributions of L. Schwartz. 
 
 }
 
 \er

\section{Acknowledgements} 
AK carried out the research within the state assignment of Ministry of Science and Higher Education of the Russian Federation for IITP RAS.


\end{document}